\documentstyle[12pt]{article}
\input psfig
\long\def\comment#1{}
\begin{document}
\title{Moving Observers in an Isotropic Universe}
\author{Subhash Kak}
\date{}
\maketitle

\begin{abstract}
We show how the anisotropy resulting
from the motion of an observer in an isotropic universe
may be determined by measurements.
This provides a means to identify inertial frames, yielding
a simple resolution to the twins paradox of
relativity theory.
We propose that isotropy is a requirement for a frame
to be inertial; this makes it possible to relate motion to
the large scale structure of the universe.


\end{abstract}

\section{Introduction}

The twins (or clock) paradox presents a central problem related to
the interpretation and use
of the relativity principle, namely how to
identify inertial frames. The problem concerns two twins, one of
whom leaves the earth on a spacecraft moving
with a speed comparable to the speed of light. After quick
acceleration, the craft moves in uniform motion to a distant star, swings
around and returns to the earth. When the traveling twin has returned
home, he finds he
is younger to the twin who remained earth-bound even though with respect
to him, it is the earth-bound twin who had been in motion.

Excepting for brief intervals when the traveling twin
was accelerating, the twins, according to the relativity theory, belong to
inertial frames and, therefore, the situation appears perfectly
symmetrical.

There exist many different ``resolutions'' to
the paradox. The most common of these invokes asymmetry as the
twin who leaves the earth undergoes acceleration whereas the
earthbound twin does not. Another explanation is based on 
the asymmetry in the Doppler-shifting
of light pulses received by the twins from each other in the
outward and inward journeys. Yet another
explains the age difference
to the switching of the inertial frames by the traveling twin.
These various ``resolutions'' are not
 in consonance with each other. 

The slowing down of all clocks and processes -- including atomic
vibrations -- on the traveling twin
 cannot be laid on the periods of acceleration
and turning around during the journey, since they can, in principle, be made as
small as one desires. 
Furthermore, if there is a slowing down of the clock of
the traveling twin on the complete trip, a component of this slowing
down must have occurred on the outbound trip itself, making the
paradox even more acute.

Einstein's own
``resolution'' in 1918, which was an attempt to counter the criticism
related to the paradox until that time [1], 
used the gravitational time dilation of the 
theory of general relativity to explain the 
asymmetrical time dilation of the traveling twin. This
explanation is generally considered wrong (see e.g. [2]), and
is different from the other resolutions recounted earlier. 

The diversity and the mutual inconsistency of the offered
solutions only reinforces the reality of the 
paradox within relativity. 
According to the recent assessment by Unnikrishnan [2], 
`` The failure of the accepted views and resolutions can be traced to the fact 
that the special relativity principle formulated originally for physics in 
empty space is not valid in the matter-filled universe.'' 

In this article, we present a new principle for the identification
of inertial frames in a matter-filled universe that allows us to
easily resolve the twins paradox.
The principle implies that the identification of a frame
as being inertial depends on whether the  universe has
spatial isotropy with respect to it.
This is equivalent to determining the
motion of objects against the background of distant stars.

\section{Laws and the nature of the universe}
According to the principle of relativity, systems of reference
moving uniformly and rectilinearly with respect to each other have the
same laws, and the speed of light is constant in 
all such systems. ``Laws'' are defined in operational terms, by means of
readings on instruments that are to be used in a clearly
specified manner in all inertial reference frames, in which free particles 
move in straight lines.

Poincar\'{e} enunciated the principle of relativity 
in 1904 in following words [3], [4]:
\begin{enumerate}
\item 
The
laws of physical phenomena must be the same for a `fixed' observer
as for an observer who has a uniform motion of translation relative to him:
so that we have not, and cannot possibly have, any means of discerning whether
we are, or are not, carried along in such a motion. 
\item
From all these results there must arise an entirely new kind of dynamics, {\it
which will be characterised above all by the rule, that no velocity can 
exceed the velocity of light.}
\end{enumerate}

Neither Poincar\'{e} nor Einstein, with his similar statement of the relativity
principle [5], considered its implications for determining the physical nature
of the universe. 

Before the
advent of the principle of relativity, it was popular
to conceive of the universe as being suspended in absolute space.
Given the unstated assumption of a finite
universe, it led to the notion that the centre of gravity of the universe
should be considered to be absolutely at rest, and the plane in which
the angular momentum of the universe around this centre is the greatest,
should also be considered to be absolutely at rest.

In the late 19th century, the success of the wave theory of light
spurred scientists to determine the
earth's motion relative to the aether that was taken to be the
medium in which the waves propagated.
But the failure of the efforts to measure this velocity
led Poincar\'{e} in 1899 to declare that ``absolute motion
is indetectible, whether by dynamical, optical, or electrical means.'' [6]

It seems reasonable to assume
that physical processes are a consequence of the large 
scale nature of the universe as is clear from the Coriolis
forces that tell us that the earth is rotating with respect to
distant stars. Likewise, the earth's rotation
is inferred by the fact that it is slightly flattened at the poles due
to the centrifugal forces.

Since the imperative in science is to take the
laws to be the same everywhere, the universe must
be isotropic.
Expressed differently,
the universal application of the principle of relativity 
is a consequence of the fact
that the world is isotropic.

By extension, if the universe deviated from isotropy and if its structure
was different in the past, the constants of physical laws (or
perhaps the laws themselves)
will vary with respect to location and time. Indeed, there are current
theories that propose variation of speed of light as well
as variation in the fine structure constant in the past.

If the laws are independent of the size and the structure of the visible universe
then this universe may be infinite in extent.
In such a case, all inertial frames in mutual uniform motion must be 
equivalent, as supposed by Poincar\'{e} and Einstein.

\section{Distribution of speeds}
Consider an isotropic universe in which objects are receding from 
the observer with speeds that vary uniformly over $(-1,1)$, where the speed of light,
$c$, is taken to be 1. The observer can make measurements of Doppler
shifts and conclude
that the probability density function of the speeds, $x$, is uniform:

\begin{equation}
f_X (x) = \frac{1}{2}, ~-1 \leq x \leq 1
\end{equation}

Now, suppose the observer starts moving in a specific direction
with speed $v$. The probability density function of the speeds of
the objects with respect to the observer would be changed.

The observer can measure the Doppler with respect to the distant
receding objects in the antipodal directions related to the motion.
Let the new variable of speed be $y$. By the law of composition of
velocities in relativity theory, the new speed will be given by the equation:
\begin{equation}
y = \frac{x + v}{1+ xv}
\end{equation}

Since the mapping between $x$ and $y$ is a monotonic function, one
can determine the probability density function of the variable $y$
by a simple transformation rule on random variables.
First, we compute the derivative of the mapping between $x$ and $y$:

\begin{equation}
\frac{d y}{d x} = \frac{1- v^2}{(1+ xv)^2}
\end{equation}

This implies that the new probability density function is:

\begin{equation}
f_Y (y) = f_X (x) \left|\frac{d x}{d y} \right| = \frac{1}{2} \frac{(1+ xv)^2}{(1- v^2 )}
\end{equation}

Written in terms of the variables $y$ and $v$ alone, we have
\begin{equation}
f_Y (y) = \frac{1}{2} \frac{(1- v^2 )}{(1- y v)^2 } , ~-1 \leq y \leq 1
\end{equation}

Thus the speed of the observer may be inferred in principle by measurements
of the distribution of speeds of the receding distant objects in the 
direction of the motion.

\vspace{0.2in}
\noindent
{\bf Example.} If $v = 0.9$, that is if the observer moves with a speed
that is 90 percent of the speed of light, the distribution of the recessional
speeds of the distant stars in the direction of the motion will be given by
\begin{equation}
= \frac{1}{2} \frac{0.19}{(1-0.9 y)^2} , ~-1 \leq y \leq 1
\end{equation}

\begin{figure}
\hspace*{0.2in}\centering{
\psfig{file=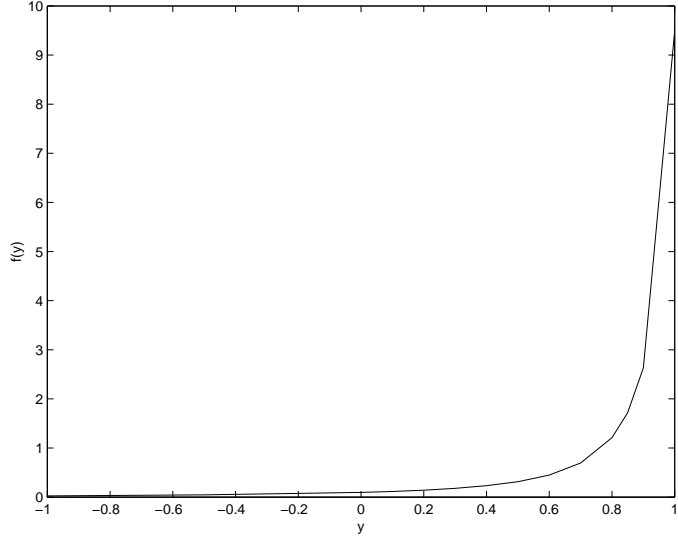,width=9cm}}
\caption{Probability density function of new speeds for $v$=0.9}
\end{figure}

Figure 1 presents the probability distribution function for $y$ for this
case of $v=0.9$.  
The anisotropy in the directions of motion is great.
To such an observer, more distant objects in the direction of the
motion would appear to recede at speeds that are close to the speed of light.
Furthermore, there is marked anisotropy with respect to the antipodal direction.

To see the values at the extreme points, we first consider $y=1$. 
\begin{equation}
f(y=1) = \frac{1}{2} \frac{(1+v)}{(1-v)}
\end{equation}

At the other end, when $y = -1$,
\begin{equation}
f(y=-1) = \frac{1}{2} \frac{(1-v)}{(1+v)}
\end{equation}

\noindent
In other words, the anisotropy is determined most clearly by considering objects 
that are furthest to the observer.

In principle, it is not essential for the speeds distribution to have been
uniform in the beginning to estimate a departure from it.

\section{Anisotropy and flow of time}
Consider an observer, A, who is at rest at origin O of an inertial frame. 
Observer, B, is also at rest with respect to the same frame. 
Now B quickly accelerates and then moves off at uniform velocity
and after reaching a point C, he reverses his direction and returns to O
with the same velocity. Both observers carry ideal clocks. 
The clocks of A and B will show local time as $T_A$ and $T_B$, where

\begin{equation}
T_A = \frac{T_B}{\sqrt{(1-\frac{v^2}{c^2})}}
\end{equation}

If $v \ll c$, one can write this equation as 

\begin{equation}
T_A \approx T_B (1 + \frac{v^2}{2 c^2})
\end{equation}

In other words, if A and B are twins,
and $v \ll c$, then twin B would be younger roughly by the amount:

\begin{equation}
T_A -T_B \approx \frac{v^2}{2 c^2} \times T_B
\end{equation}

Because of the symmetry of the travel, 
a slowing down must occur in both the outbound and the inbound
legs of the journey.
Therefore,
the slowing down must
be the consequence of a real difference between the two frames
with respect to the rest of the universe.

We propose that the difference in aging is a function, $g$, of the
two probability density functions, where we now use the same index set $x$. Or,
\begin{equation}
T_A -T_B = g( f_X (x) , f_Y (x) )
\end{equation}

If the difference between the two times is a logarithmic
function of the ratios of the probability density functions, it
ensures that the difference is zero for two frames that have
no relative motion: 

\begin{equation}
T_A -T_B = g( \log (f_X (x) / f_Y (x) ))
\end{equation}

One might also use a relative entropy measure to compare the two distributions:

\begin{equation}
H(X:Y) =  \sum f_X (x) \log \frac{f_X (x)}{f_Y (x)}
\end{equation}

Having seen that
the uniform motion with respect to the distant stars 
is detectable, 
it is possible to determine which of the two twins is to be
taken to be inertial. 
This means that the slowing down of clocks is a consequence of the
universe not being perfectly isotropic to it.

Our analysis shows that Poincar\'{e}'s and Einstein's implicit 
equivalence of all frames
in uniform motion in the principle of relativity is incorrect in a finite
universe.

Time dilation is seen as a consequence of Lorentz transformations, but it
may also be viewed 
as a consequence 
of the anisotropic apprehension of the universe by the moving
observer.
This allows easy resolution of various
situations related to observers.

For example, if two travelers leave in arbitrary directions with the same
speed and return to an inertial frame, their own clocks would still be
synchronized, but lag behind that of the inertial frame.


One could also speak of a set of frames that can be put in an hierarchical
relationship in terms of their suitability as being inertial.

\section{Concluding remarks}
If it is assumed that ``physical laws'' are a consequence of the
large scale  nature of the universe, then it is valid to assume
that there will be a difference in the experience of two observers in relative
uniform motion 
if isotropy of the universe is not maintained by them equally.

Could one say that the assumption of a finite universe goes against
the relativity principle as normally defined
because in such a universe it is possible, in
principle, to calculate the distribution of speeds?
If the universe is infinite in size then
the change in its isotropy with respect to the moving observer may not
materially alter the isotropy maintained with the non-visible part of the
universe.
In such a case, Poincar\'{e}'s and Einstein's claims regarding the
equivalence of all frames in relative uniform motion remain valid, but 
at the cost of the twins paradox.

It must be stressed that the determination that an observer is in motion with
respect to distant stars does not imply a return to absolute space. An isotropic
universe will have an infinity of inertial frames.

\end{document}